\documentstyle[12pt]{article}
\def \o  {\overline}
\def \ra {\rightarrow}
\def \f  {\frac}
\def \non {\nonumber}

\def \lc {{\cal L}}
\def \ba {\begin{array}}
\def \ea {\end{array}}
\def \be {\begin{equation}}
\def \ee {\end{equation}}
\def \bea{\begin{eqnarray}}
\def \eea{\end{eqnarray}}

\def \al{\alpha}
\def \bt{\beta_t}
\def \bg{\beta_{\gamma}}

\def \om {\omega}

\def \la{\lambda}

\def \lg {\lambda_{\g}}

\def \lt {\lambda_t}

\def \tl{\theta_l}
\def \tht{\theta_t}
\def \th0{\theta_0}

\def \zgg{$Z\gamma \gamma$}
\def \zzg{$ZZ\gamma$}

\def \l  {Lagrangian}

\def \cp {$\boldmath CP$}

\def \g {\gamma}

\def \gg {\g \g}

\def \tbar {$\o t$}

\def \ttbar{$t \o t$}

\def \ep {$e^+ e^-$}

\def \ggtt {$\g \g \ra t \o t$}

\begin{document}
\begin{flushright}
	   PRL-TH/98-10\\
           TIFR/TH/98-34\\
           August, 1998\\
           hep-ph/9809203\\[3mm]
\end{flushright}

\begin{center}
{\large \bf 
\cp-violating \zgg\ and top-quark 
electric dipole couplings in \ggtt}\\[1cm]
P. Poulose$^{\bf a,}$\footnote{poulose@theory.tifr.res.in}
and Saurabh D. Rindani$^{\bf b,}$\footnote{saurabh@prl.ernet.in}\\[5mm]
$^{\bf a.}$ {\it Dept. of Theoretical Physics, Tata Institute of Fundamental
Research,}\\ {\it Mumbai-400 005, India}\\[2mm]
$^{\bf b.}$ {\it Theory Group, Physical Research Laboratory,}\\
{\it Navrangpura, Ahmedabad-380 009, India}\\[15mm]
\end{center}

\abstract{
An effective anomalous \cp-violating \zgg\ coupling can give rise to observable
\cp-odd effects in \ggtt. We study certain asymmetries in the decay lepton
distributions in \ggtt\ arising from top decay in the presence of a
\cp-violating \zgg\ coupling as well as a
top-quark electric dipole coupling. We find that
a photon linear collider with geometric luminosity of 20 fb$^{-1}$ 
can put limits of the order of 0.1 on the imaginary part of the \cp-violating 
anomalous \zgg\ coupling using these asymmetries.
}\\[1cm]

While gauge-boson couplings to fermions have been measured with great
accuracy and agreement of these measurements with predictions from
the Standard Model (SM)
is overwhelmingly precise,
the area of  pure gauge bososn couplings is not explored with that
precision.  Deviation of the gauge boson couplings from the SM values
could be used to infer the  presence of new physics. 
Such couplings arising from 
new physics could even be \cp-violating. 
There have
been detailed discussions on the anomalous triple gauge
boson couplings $WWV$ and $Z\gamma V$, where $V=\gamma,
Z$, in the literature \cite{wwv,zgg,sdr}.  Experimental bounds 
on these couplings obtained at LEP \cite{aleph,delphi,l3,opal} and at the
Tevatron \cite{cdf,d0} are fairly weak, and are found to be consistent 
with SM. While future experiments would improve these limits, effects of 
these couplings are expected to be more visible at higher energies, for
example, at the proposed linear $e^+e^-$ colliders. 
\cp-violating triple gauge boson couplings get contributions only beyond 
one loop in SM. This makes them good candidates to study new physics
effects. 

Theoretical studies have largely concentrated on $\gamma W^+W^-$ and
$ZW^+W^-$ couplings, and less attention has been paid to neutral gauge-boson
self-couplings. In particular, \cp-violating \zgg\ and \zzg\ couplings have been
the subject of few discussions. These couplings are absent at the tree level
in SM, and any observation of these at a substantial level would signal new
physics beyond SM. Our work concerns the measurement 
of the \cp-violating \zgg\ coupling.

Most experimental studies of anomalous gauge boson couplings at $e^+e^-$ or
hadron colliders would have to
deal with the problem of separating not only several different types of form
factors for the same effective vertex, but also separating couplings involving
\zgg\ couplings from \zzg\ couplings. In this respect, 
projected $\gg$ colliders would have an advantage that with the initial state
fixed as $\gg$, only the \zgg\ vertex would contribute, ignoring triple-photon
couplings. For this reason we would like to advocate here the use of a
$\gg$ collider for a study of \zgg\ couplings.   

With these points in mind we discuss in this letter the 
effect of \cp-violating \zgg\ coupling in the process \ggtt. 
The top quark is expected to decay before it hadronizes \cite{bigi},
and therefore one has
the hope of using decay correlations to deduce polarization information of the
production process.
In this process, however, one
has to contend with a possible extra source of \cp\ violation, viz., 
the \cp-violating electric dipole coupling of the top quark. We have also
discussed here this possibility, and ways of obtaining separate limits on the
\cp-violating \zgg\ and top dipole couplings.

Photon linear colliders  have been widely  discussed   in the
literature.
In such a collider an intense low-energy laser beam would
be scattered in the backward direction 
by a high-energy electron beam, transferring most of
the electron energy to the photon in the process.
The photon beam thus produced is
made to collide  with another photon  beam produced in a similar  way.
The main  features of such  a  photon linear collider  are described in
Ref.  \cite{ginz}. The luminosity and  the polarizations of the photon
beams would depend on the initial electron and laser beam helicities as
well as their energies.  
When the electron and the laser beam helicities are of the opposite sign, 
the photon spectrum peaks at higher energies. Also, in this case, the higher
energy photons will have the same helicity as that of the initial electrons
\cite {ginz}.
But in general the scattered photon will be in a mixed polarization
state. As we shall see, polarization plays an important role in improving the
sensitivity of the experiments we suggest. Here we concentrate on
longitudinally polarized electron beams and circularly polarized laser beams.

For the purpose of studying the anomalous  \cp-violating \zgg\ coupling, 
we consider the
effective \l\ 
\bea
\lc_{\it eff}&=&\lc_{SM}+\lc_{AC},
\label{eq:efflag}
\eea
where $\lc_{SM}$ is the usual SM Lagrangian, and
\bea
\lc_{AC}&=&\f{e}{16M_Z^2\cos\theta_w\sin\theta_w}
	\,\left[\la_1\,F_{\mu \nu}\:F^{\nu \lambda}\:
	(\partial_\lambda Z^\mu+\partial^\mu Z_\lambda)
+ \la_2\,F_{\mu\nu}\,F^{\mu\nu}\,\partial^{\lambda}Z_{\lambda}\right],
\eea
with 
\bea
F_{\mu \nu}=\partial_\mu A_\nu-\partial_\nu A_\mu,
\eea
$A_\mu$ and $Z_\mu$ being the photon and $Z$-boson fields, and $\lambda_1$ and
$\la_2$ are 
dimensionless couplings. 
$\lc_{AC}$ is the most general \cp-violating \l\ consistent with Lorentz and 
electromagnetic gauge
invariance, if the photons are on-shell, or coupled to conserved currents.
In our case, the photons are on-shell. The second term in $\lc_{AC}$ is absent 
if
the $Z$ is on-shell or is coupled to a conserved current. The \cp-violating
\l\ with only the first term was used in \cite{sdr} for calculating
\cp-violating forward-backward asymmetry in $e^+e^-\rightarrow \gamma Z$.
It would be possible to include terms with more derivatives on the
fields, but these can be taken care of by assuming that $\lambda_{1,2}$ are not
 just 
constants, but in momentum space, they are form factors depending on invariants
constructed out of momenta.

We now derive the \zgg\ vertex arising from the above effective \l\
$\lc_{AC}$.  It turns
out that the contributions of the two terms in $\lc_{AC}$ are proportional to
each other, and the final vertex 
$ie \Gamma^{AC}_{\al\beta\mu}$ can be written as
\bea
\Gamma^{AC}_{\al\beta\mu}(k_1,k_2,q)&=&
\f{i\la}{16\,M_Z^2\,\cos\theta_w \sin\theta_w}\non\\
&&\times \left[ \right.
g_{\alpha \nu}\:(k_2.q\:k_{1\beta}-k_1.k_2\:q_\beta)\non\\
&&+g_{\beta \nu}\:(k_1.q\:k_{2\alpha}-k_1.k_2\:q_\alpha)\non\\
&&-g_{\alpha \beta}\:(k_2.q\:k_{1\nu}+k_1.q\:k_{2\nu})\non\\
&&+k_{2\alpha} q_\beta k_{1\nu}+k_{2\nu} q_\alpha k_{1\beta}
\left. \right]
\eea
where $k_1,\;k_2$ and $q$ are the four momenta of the photons and
the $Z$ boson, and $\al,\;\beta$ and $\mu$ are the corresponding 
Lorentz indices. Now $\la$ is a linear combination (albeit momentum dependent)
of $\la_1$ and $\la_2$. We will henceforth discuss constraints only on this
combined form factor $\la$.

The process $\gamma\gamma\rightarrow t \overline{t}$ gets contribution, apart
from the standard model $t$ exchange diagrams, also from the anomalous \zgg\
vertex, with a virtual $Z$ exchanged in the $s$ channel. (We neglect a
possible $\gamma\gamma\gamma$ vertex).
Using the method discussed by Vega and Wudka \cite{vega} we compute the
helicity amplitudes for the process \ggtt. It turns out that the
\cp-violating \zgg\ coupling contributes only  when both 
of the photon beams have the same  helicity, as well as the
top quark and the top antiquark have the same helicity. The amplitude in
this case is given below, including the effect of the 
top-quark electric dipole form factor (EDFF).
The EDFF occurs in the \l\ term,
\bea
\lc_{\it edff}&=&ie\,d_t\,\o{\psi}_t\,\sigma^{\mu \nu}\,
\gamma_{\small 5}\,\psi_t\;F_{\mu \nu}, \label{eq:ldt}
\eea
and its effects were 
discussed in the earlier work \cite{pp1,choi}. 

The helicity
amplitudes for the process \ggtt\ with these two \cp-violating couplings, for
the case when the two photons have equal helicities, as do the $t$ and
$\overline{t}$, are given by
\bea
M(\lg,~\lg,\lt,~\lt)&=& -\f{4 m_t\,e^2\,Q_t^2}{\sqrt{s}
        (1-\bt^2\cos^2\theta_t)}
        \left\{(\lg+\lt \bt) \right. \non \\
        &&\left.-i\,d_t\;2m_t\left[2+\f{s}{4m_t^2}\bt
        (\bt-\lt \lg) \sin^2\theta_t\right]\right\}\non\\
	&&+ i e^2\,\la\;\f{m_t}{8\sqrt{s}x_w(1-x_w)} 
	\left(\f{s}{4m_Z^2}\right)^2,
\eea
where $\lg,~\lg',\lt,~\la_{\o t}$ in $M(\lg,~\lg',\lt,~\la_{\o t})$
are the helicities of the two photon beams and
the top quark and the top antiquark. $\sqrt{s}$ is the c. m. energy,
$m_t$ and $m_Z$ are the top quark
and Z boson masses, $Q_t$ is the electric charge of the top quark, 
$x_w$ is given in terms of the weak mixing angle $\theta_w$ by
$x_w=\sin^2\theta_w$, and
$\bt$ and $\theta_t$ are the velocity and
the scattering angle of the top quark in the c.m. frame.
We have dropped terms quadratic in $d_t$.
Rest of the amplitudes do not depend on the \zgg\ coupling and are given 
in Ref. \cite{pp1}. It is interesting to note that in the last term in the
amplitude, not only is the factor of $(s-m_Z)^{-2}$ from the $Z$ propagator
cancelled by a factor coming from the anomalous vertex, but there is an
additional $s^{3/2}$ dependence which increases with energy.

We construct \cp-violating asymmetries which can be used to study the 
effect of the new
coupling in experiments. In principle, there would be definite predictions for
top and antitop polarizations in the presence of $CP$-violating terms. These
could be used to isolate $CP$ violation. In particular, the anomalous coupling
contributes only to the amplitude with equal $t$ and $\overline{t}$
helicities. The interference of this amplitude with the standard model
amplitude can give rise to definite prediction for the polarizations. However,
it is the final top decay products which have to be used to analyze the
polarization. From a practical point of view, it is better to work with
asymmetries in terms of the decay leptons. 

The asymmetries discussed below 
were studied
earlier \cite{pp1} in the context of \cp\ violation effects induced by a
possible
top-quark electric dipole form factor (EDFF).
These asymmetries are (i) the asymmetry in the number
of leptons and the antileptons produced as decay products of the top quark
and the top antiquark (the charge asymmetry) and (ii) the sum of the
forward-backward asymmetries of the leptons and the antileptons.   Being
independent of the top quark momentum these asymmetries are experimentally
favourable. 

The two asymmetries are written in terms of the differential cross section as
follows.
\be
A_{ch}(\theta_0)=\frac{
{\displaystyle          \int_{\theta_0}^{\pi-\theta_0}}d\theta_l
{\displaystyle          \left( \frac{d\sigma^+}{d\theta_l}
                -       \frac{d\sigma^-}{d\theta_l}\right)}}
{
{\displaystyle          \int_{\theta_0}^{\pi-\theta_0}}d\theta_l
{\displaystyle          \left( \frac{d\sigma^+}{d\theta_l} +
\frac{d\sigma^-}{d\theta_l}\right)}}
\label{eq:ach}
\ee
and
\be
A_{fb}(\theta_0)= \frac{ {\displaystyle
\int_{\theta_0}^{\frac{\pi}{2}}}d\theta_l {\displaystyle
\left( \frac{d\sigma^+}{d\theta_l} +
\frac{d\sigma^-}{d\theta_l}\right)} {\displaystyle
-\int^{\pi-\theta_0}_{\frac{\pi}{2}}}d\theta_l {\displaystyle
\left( \frac{d\sigma^+}{d\theta_l} +    \frac{d\sigma^-}{d\theta_l}
\right)}}
{
{\displaystyle          \int_{\theta_0}^{\pi-\theta_0}}d\theta_l
{\displaystyle          \left( \frac{d\sigma^+}{d\theta_l} +
\frac{d\sigma^-}{d\theta_l}\right)}}.
\label{eq:achfb}
\ee
In the above equations, $\frac{d\sigma^+}{d\theta_l}$ and
$\frac{d\sigma^-}{d\theta_l}$ refer respectively to
the $l^+$ and $l^-$ distributions in the c.m. frame of the $\gg$ pair,
$\theta_l$ is the polar angle of $l^+$ or $l^-$ 
and $\theta_0$ is the cut-off in $\theta_l$ in the forward and backward
directions. Not only is a cut-off in $\theta_l$ necessary from the experimental
detection point of view, it also helps to tune the sensitivity of the
experiments, as discussed below.

The asymmetries discussed above being $CPT$-odd should be
proportional to the absorptive part of the amplitude.\footnote{Here $T$ 
refers to the
naive time-reversal operator, which reverses spins and momenta of the particles
involved, while not interchanging the initial and final states.}

Photon-photon collisions would be achieved at an \ep\ linear collider. So the
actual collision rate for \ep\ into a given final state is a convolution of
the two-photon collision rate into that final state with the spectra of photons
from laser backscattering. Denoting the effective two-photon luminosity by
$L_{\g\g}$, we use the expression for the differential luminosity
$d\,L_{\g\g}$ derived in \cite{ginz}. We refer the reader to this work for
details.
Since it is easier to calculate the differential cross section in the
center of mass (c.m.) frame of the $\gg$ pair, we modify the above
expression changing the limits to take care of the boost needed to go to 
the lab frame from the $\gg$ c.m. frame. We can then write
\bea
A_{ch}&=&\f{1}{2\,N}\left\{\int \f{d L_{\g
\g}}{d\om_1\,d\om_2}\,d\om_1\,d\om_2
        \int_{-1}^1\!d\cos\tht\right.\non \\
        &&\times\left.
        \int^{f(\theta_0)}_{g(\theta_0)}
        d\cos\tl\left[\frac{d\sigma^+}{d\cos\tht\,d\cos\tl}-
      \frac{d\sigma^-}{d\cos\tht\,d\cos\tl}\right]\right\},
\label{eq:ggach2}
\eea
and
\bea
A_{fb}&=&\f{1}{2\,N}\,\int \f{d L_{\g \g}}{d\om_1\,d\om_2}\,d\om_1\,d\om_2
        \int_{-1}^1\!d\cos\tht\non \\
        &&\times \left\{
        \int^{f(\theta_0)}_{-\beta_\g}
        d\cos\tl\left[ \frac{d\sigma^+}{d\cos\tht\,d\cos\tl}+
      \frac{d\sigma^-}{d\cos\tht\,d\cos\tl}\right]\right.\non \\
        &&\left.-
        \int^{-\beta_\g}_{g(\theta_0)}
        d\cos\tl\left[ \frac{d\sigma^+}{d\cos\tht\,d\cos\tl}+
      \frac{d\sigma^-}{d\cos\tht\,d\cos\tl}\right]\right\}.
\label{eq:ggafb}
\eea
Here 
\[
f(\theta_0)=\f{\cos\theta^{cm}_0-\bg}{1-\bg \cos\theta^{cm}_0}
\] and
\bea
g(\theta_0)&=&\f{\cos(\pi-\theta^{cm}_0)-\bg}{1-\bg
        \cos(\pi-\theta^{cm}_0)} \non \\
&=&\f{-\cos\theta^{cm}_0-\bg}{1+\bg \cos\theta^{cm}_0}, \non
\eea
$\om_1$ and $\om_2$ are the two photon beam energies in the lab frame,
$\theta_t$ is the scattering angle of the top quark in the c.m. frame of
the \ttbar\ system, $\theta_0^{cm}$ is the cut-off in this c.m. frame,
and $\bg=(\omega_1 -\omega_2)/(\omega_1 +\omega_2)$.
$N$ is the total number of events produced. 

It is most advantageous to make use of the 
the semileptonic decay of \ttbar\ pair, wherein either
of $t$ or \tbar\ decays leptonically, while the other decays hadronically.
While hadronic decays have large branching ratios, purely hadronic events are
difficult to detect because of the large background.
On the other hand, branching ratio for decay into leptons is small, even though
the signal is clear. With a semileptonic final state, the overall branching
ratio is still not too low, and since our asymmetries involve measurement only
on a single lepton, the signal is easily measureable.
In our calculations, all integrations except
the $\om_1$, $\om_2$ and $\theta_t$ are done analytically. These three 
integrations are done numerically. 

Sensitivity of the measurement of these asymmetries at specific colliders
can be calculated considering the statistical fluctuations. 
For the asymmetry to be observable at the 90\% confidence
level (C.L.), the number of
asymmetric events should be larger than 
$1.64\:\sqrt{N}$, where $N$ is the total number of semi-leptonic events
produced.
This means that the asymmetry has to have a minimum value of 
\bea
A_{min}=\f{1.64}{2\sqrt{N}}.
\eea
We have kept only the linear terms in the anomalous coupling,
assuming that the anomalous coupling is small, and that higher-order
terms are negligible.
Thus  the asymmetry  can be written as
\bea
A=C_{ac}\:{\rm Im}\la,
\label{asy1}
\eea
where the coefficient $C_{ac}$ is independent of the coupling parameter,
$\la$.

Hence the asymmetries will be
proportional to the imaginary part of the couplings.
Thus eqn. \ref{asy1} would give a limiting value for the coupling
\bea
{\rm Im} \la \;|_{max} = \f{1.64}{2\sqrt{N}}\:\f{1}{C_{ac}}
\eea
in the case that the symmetry is not observed.

In an earlier work \cite{pp1} we had considered these same
asymmetries arising due to  the top quark electric dipole form factor
(EDFF) in the process \ggtt .
If we consider the simultaneous presence of both the 
EDFF and the \zgg\ coupling, the  asymmetry would be a function of these
two parameters. In such a case we could get limits on both these parameters
simultaneously by plotting a contour in the parameter plane. This is done
as follows.

In the presence of both these sources of $CP$ violation, the asymmetry could be written as
\bea
A=C_{ac}\:{\rm Im}\la + C_{\it edff}\:{\rm Im} \tilde{d}_t.
\eea
We have redefined the EDFF (see eqn. \ref{eq:ldt}) in terms of a dimensionless 
parameter: \(d_t=\tilde{d}_t/2\,m_t\).
$C_{ac}$ and $C_{\it edff}$ are coefficients independent of the
parameters, $\la$ and $\tilde{d}_t$.

To determine the sensitivity of the measurement, we now need the expression
corresponding to
two degrees of freedom.  For the asymmetry to be observable,
in this case, the number of asymmetric events should be greater than 
$2.15 \sqrt{N}$. This gives a linear relation between the 90\% C.L. limiting
values of the two parameters:
\bea
C_{ac}\:{\rm Im}\la^{max} + C_{\it edff}\:{\rm Im} \tilde{d}_t^{max}=\pm
\f{2.15}{2\sqrt{N}}.
\label{asy2}
\eea
The $\pm$ comes in because the asymmetry could be of either sign.
Contours plotted in the ${\rm Im}\la - {\rm Im}\tilde{d}_t$ plane would give 
a band of allowed
values of ${\rm Im}\la$ and ${\rm Im}\tilde{d}_t$ for a single asymmetry. 
Using more than one
asymmetry will give an allowed area, which is the area of intersection of the 
bands obtained for the individual asymmetries. 
Alternatively, a single asymmetry can be used with 
two different polarization combinations for the initial beams, and a similar
allowed region of intersection can be determined.

For our numerical calculations, we have assumed that the electron beams
have axial symmetry and a Gaussian distribution. We also assume that 
the conversion distance, i.e., the distance between
the conversion points of the lasers and the interaction point of the colliding
photons, is negligible. With these assumptions, as discussed in \cite{ginz},
the expressions for the
cross sections for the case with longitudinally polarized electrons and
circularly polarized laser photons simplify considerably.

We have assumed, for most of our calculations, a cut-off of at least
30$^{\circ}$ in the forward and
backward directions of the lepton momentum. This should be sufficient 
for the practical purposes of suppressing the background of forward (and
backward) moving particles due to standard-model processes. However, if a
minimum energy or transverse momentum cut-off for the detection of leptons 
is required,
our results would still be valid with such a cut-off, since most events with a
lepton energy less than about 45 GeV are found to be suppressed. 

We discuss our numerical results in the following. 

We first switch off the dipole term and consider the effect of the triple
gauge boson (\zgg) coupling. We study the asymmetries
varying the helicities of the initial
beams, the cut-off angle and the beam energy.  The
forward-backward asymmetry is seen to be more
sensitive in all cases.

Studying the asymmetries for different
combinations of the initial electron and laser beam helicities, it is seen,
as expected, that the forward-backward asymmetry is absent when both of the
electron beams have the same helicity as well as both of the laser beams have 
the same polarization. This is because, in this case, the two photon beams 
are identical and there is no distinction between the forward and the backward
directions. 

\begin{table}
\begin{center}
\begin{tabular}{rrrrrrrrr}
\hline
\\
&&&&&\multicolumn{2}{c}{Asymmetries}&\multicolumn{2}{c}{
Limits on Im $\la$ }\\
&&&&&&&\multicolumn{2}{c}{from}\\[2mm]
$\lambda_e^1$&$\lambda_e^2$&$\lambda_l^1$&$\lambda_l^2$&
N&$A_{ch}$&$A_{fb}$&$|A_{ch}|$&$|A_{fb}|$\\[2mm]
\hline
\\
$-$0.5& $-$0.5& $-$1&$-$1&55&$-$0.0031&  0.000&35.098&\\
$-$0.5& $-$0.5& $-$1&1&215&  $-$0.0049&  0.412&11.361&0.136\\
$-$0.5& $-$0.5&    1&1&631&  $-$0.0090&  0.000&3.637&\\
$-$0.5& 0.5&$-$1&$-$1& 62 &  $-$0.0035&$-$0.403&29.502&2.569\\
$-$0.5& 0.5&$-$1&   1& 23 &  $-$0.0037& 0.256 &50.354&0.661\\
$-$0.5& 0.5&   1&$-$1&163 &  $-$0.0004&$-$0.101&144.456&0.635\\[3mm]
\multicolumn{4}{c}{Unpolarized}&179&$-$0.0056&0&11.004&\\[3mm]
\hline
\end{tabular}
\end{center}
\caption{
Asymmetries and limits on the coupling obtained from
them at different helicity combinations. 
Asymmetries are for Im $\la=1$.  
Numbers are obtained assuming an integrated luminosity of
20~fb$^{-1}$. Initial electron beam energy is taken to be 
250 GeV and a cut-off angle of 30$^\circ$ is
assumed.}
\end{table}

Table 1 displays the asymmetries and the limits which can be obtained from
them for different helicity combinations for an initial electron 
beam energy of 250
GeV, a laser beam energy of 1.24 eV, an integrated geometrical luminosity of 20
fb$^{-1}$ for the \ep, and a cut-off angle $\theta_0=30^\circ$. We find that 
the best limits would be obtained from the charge
asymmetry  when both $\la_e^1=\la_e^2$ and $\la_l^1=\la_l^2$ but the product,
$2 \la_e^i \la_l^i=-1$, where
$\la_e^i$ and $\la_l^i$ are the electron and laser helicities.
We get a limit of 3.6 on Im $\la$ in this case. On the other hand the
forward-backward asymmetry gives best limits when $\la_e^1=\la_e^2$ and
$\la_l^1=-\la_l^2$. The limit obtained in this case is 0.14. We have
considered this helicity combination while studying the variation of
asymmetry with other parameters like the cut-off angle or energy.

It is clear from Table 1 that the limits that would be obtained in the absence
of polarization are poor, and the importance of using polarized beams cannot be
overemphasized.

We next consider the variation of asymmetries with the cut off angle.   The
result is shown in Table 2. As expected, there is no charge asymmetry
in the absence of a  cut-off.  This is because when there is no cut-off, the
asymmetry is just the difference in the number $t$ and \tbar, which is
zero from charge conservation.  We see from the table that the limit from
charge asymmetry is best for a cut-off around $60^\circ$, while the limit
from the forward-backward asymmetry gets better for smaller cut-off angles.

\begin{table}
\begin{center}
\begin{tabular}{rrrrrr}
\hline
&&&\\
&&\multicolumn{2}{c}{Asymmetries}&
\multicolumn{2}{c}
{Limits on Im $\la$ }\\
&&&&\multicolumn{2}{c}{
from}\\[3mm]
\multicolumn{1}{c}{$\theta_0$}&N&$A_{ch}$&$A_{fb}$&$|A_{ch}|$&$|A_{fb}|$\\[1mm]
(deg.)&&&&&\\[3mm]
\hline
&\\
  0 &249&    0.0000&    0.476& & 0.109\\
 10 &245& $-$0.0006&    0.469&  87.339   & 0.112\\
 20 &233& $-$0.0023&    0.447&  23.192  &  0.120\\
 30 &215& $-$0.0049&    0.412& 11.361   & 0.136\\
 40 &189& $-$0.0081&    0.364&   7.340   & 0.164\\
 50 &159& $-$0.0115&    0.305&   5.668   & 0.213\\
 60 &123& $-$0.0146&    0.237&   5.049   & 0.311\\
 70 &84&  $-$0.0172 &   0.162&   5.202   & 0.551\\
 80 &42& $-$0.0188   & 0.082&   6.660   & 1.525\\[3mm]
\hline
\end{tabular}
\caption{
Variation of asymmetries and limits on the couplings 
obtained from them with cut-off angle. 
Asymmetries are for Im $\la=1$.  
Numbers are obtained assuming an integrated luminosity of
20~fb$^{-1}$. Initial electron beam energy is taken to
be
250 GeV and the helicity combination considered is
$\la^1_e=-0.5,\;\la^2_e=-0.5,\;\la^1_l=-1$ and
$\la^2_l=1$.} 
\end{center}
\end{table}

Asymmetries are best for higher \(x=4E_{\rm b}\om_0/m_e^2\) 
($\om_0$ is the laser beam energy and $E_{\rm
b}$, the electron beam energy) values. However, there is a limit to which 
the $x$ value can be increased.
For $x>4.83\,$ \ep\ production due to the collision of high energy
photon beam with laser beam is considerable \cite{ginz}.
This introduces additional \ep\ beam backgrounds as well as degrading the
photon spectrum.
We use a value of 4.75 for $x$. 
With higher $E_{\rm b}$ the sensitivity increases considerably upto a point,
and then increases more slowly. The improvement is by an order of magnitude in 
going from $E_{\rm b}=250$ GeV to $E_{\rm b}=500$ GeV. Table 3 displays the
values obtained with varying electron beam energy.

\begin{table}
\begin{center}
\begin{tabular}{rrrrrr}
\hline
&&&\\
&&\multicolumn{2}{c}{Asymmetries}&
\multicolumn{2}{c}
{Limits on Im $\la$}\\
&&&&\multicolumn{2}{c}{
from}\\[3mm]
\multicolumn{1}{c}{$E_{\rm b}$}&N&$A_{ch}$&$A_{fb}$&$|A_{ch}|$&$|A_{fb}|$\\[1mm]
(GeV)&&&&&\\[3mm]
\hline
&&&\\
250&   215&   $-$0.005  &  0.412 & 11.361  &  0.136\\
500&  1229&   $-$0.348  &  3.914  & 0.067  &  0.006\\
750&  1032&   $-$1.087  &  6.695  & 0.024  &  0.004\\
1000&  850&   $-$1.879  &  8.142  & 0.015  &  0.004\\[3mm]
\hline
\end{tabular}
\caption{
Asymmetries and the limits at different electron beam
energies. Cut off is taken to be $30^\circ$ and a
helicity combination of 
$\la^1_e=-0.5,\;\la^2_e=-0.5,\;\la^1_l=-1$ and
$\la^2_l=1$ is considered. 
Asymmetries are for Im $\la=1$.  
An integrated luminosity of 20~fb$^{-1}$ is assumed.}
\end{center}
\end{table}
\begin{figure}[ht]
\vskip 8.5cm
\includegraphics{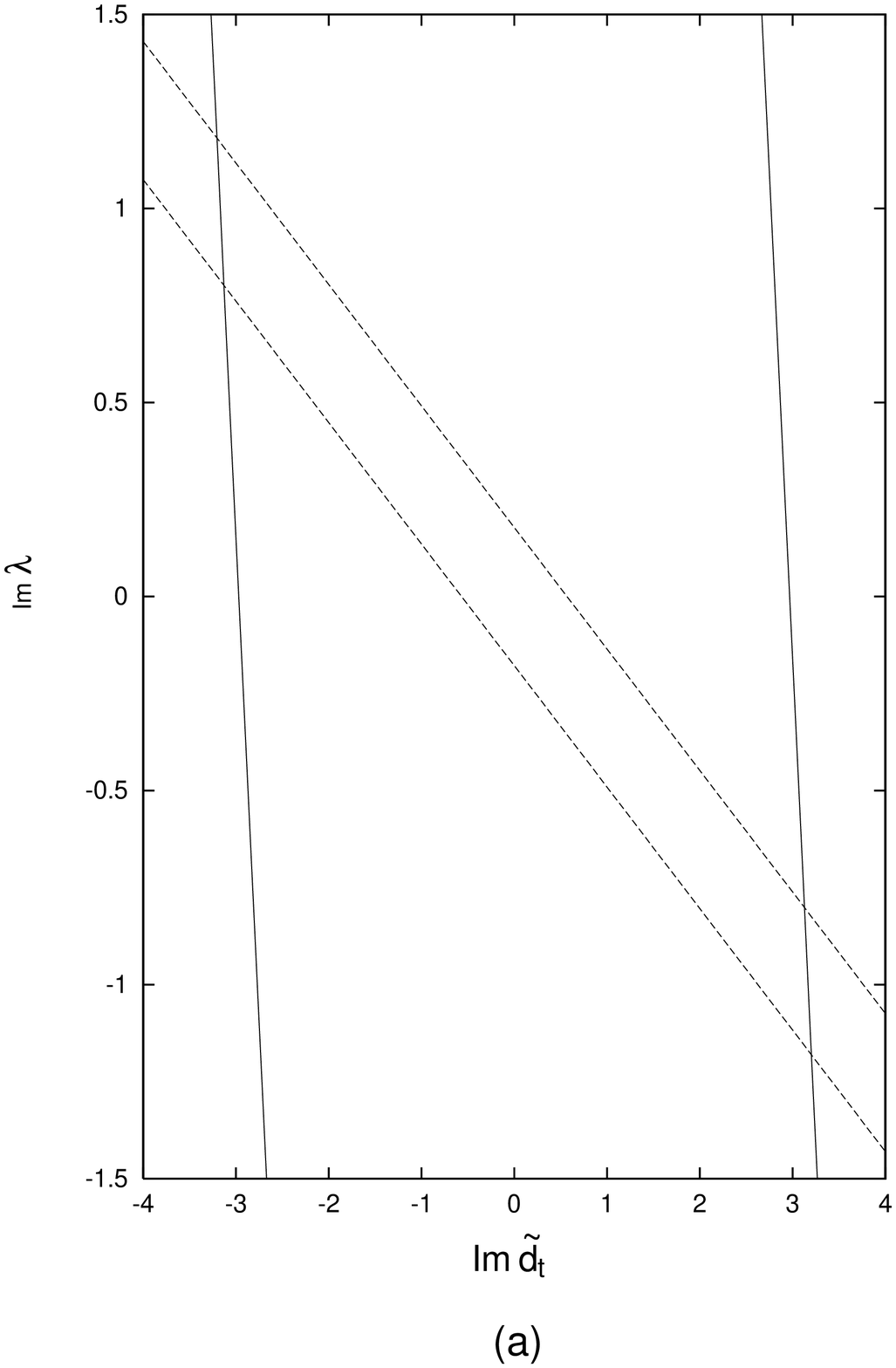}
\includegraphics{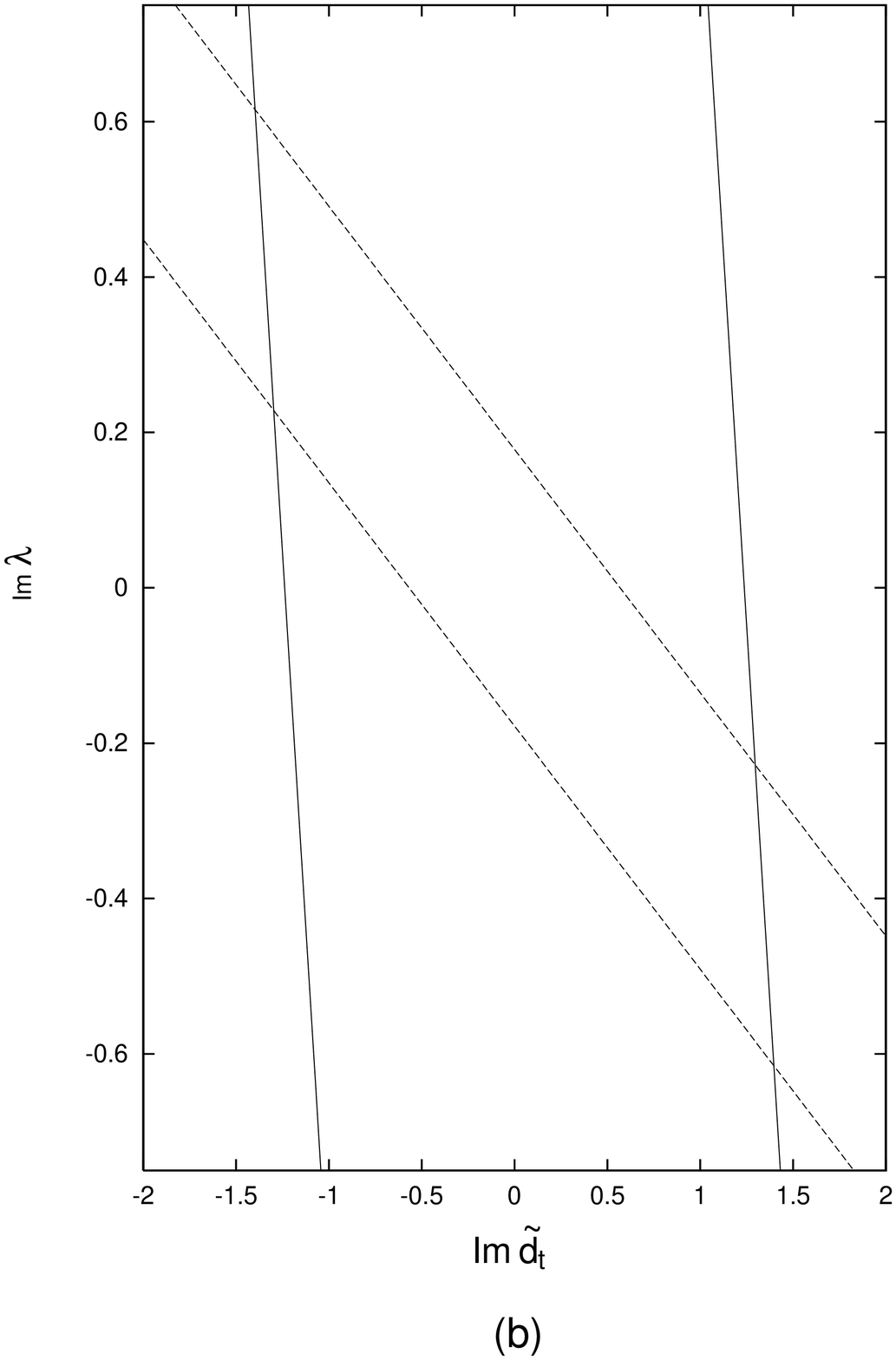}
\caption{ 
(a) Contours in the Im$\lambda - {\rm Im} \tilde{d}_t$ plane obtained from the 
charge
asymmetry (solid lines)
and forward-backward asymmetry (dashed lines) with helicity combination
$\lambda_e^1=\lambda_e^2=-0.5,\;\;\lambda_l^1=-\lambda_l^2=-1$.
(b) Contours in the Im$\lambda - {\rm Im} \tilde{d}_t$ plane obtained from the
charge asymmetry with the 
helicity combination $\lambda_e^1=\lambda_e^2=-0.5,\;\;
\lambda_l^1=\lambda_l^2=1$ (solid lines) and forward-backward asymmetry with
the helicity combination 
$\lambda_e^1=\lambda_e^2=-0.5,\;\;
\lambda_l^1=-\lambda_l^2=-1$ (dashed lines).
A cut-off $\theta_0=30^{\circ}$, initial electron beam energy
E$_{\rm b}=250$ GeV, and an integrated geometric luminosity of the
20~fb$^{-1}$ are assumed.
}
\end{figure}

Considering the case where both the EDFF and the triple gauge boson
coupling $\la$ are present, we get simultaneous limits on these
parameters by plotting contours in the ${\rm Im}\la-{\rm Im}\tilde{d}_t$ plane.
We get a band of allowed region in the parameter space for each asymmetry
by using eqn.\ref{asy2}. By considering two asymmetries we get two bands
whose area of intersection would give the allowed values of the parameters.
With the charge asymmetry and the forward-backward
asymmetry we find that the best simultaneous limits are obtained for the 
case $\la_e^1=\la_e^2$ and $\la_l^1=-\la_l^2$. As shown in Fig 1(a)
we get a limiting value of 1.2  for Im$\la$ and $0.79\times10^{-16}$ e cm
for the EDFF, Im$d_t$. When the charge
asymmetry alone or forward-backward asymmetry alone was considered for 
different helicity combinations, the 
limits worsened. But by combining
the case of charge asymmetry for $\la_e^1=\la_e^2=-0.5,\;\;
\la_l^1=\la_l^2=1$ and forward-backward asymmetry for $\la_e^1=\la_e^2=-0.5,
\;\;\la_l^1=-\la_l^2=-1$ we get better
limits, viz., 0.6 on ${\rm Im}\la$, and $0.68\times10^{-16}\; e$ cm on ${\rm
Im}d_t$ (Fig 1(b)). 

Some remarks about the magnitudes of the limits on $\lambda$ are in order.
$\vert \lambda \vert$ would be bounded by unitarity. However, the corresponding
limits are quite weak upto a fairly high energy scale \cite{sdr}.

The electron electric dipole moment (EDM) has been measured
with very high precision and the experimental value has been
presented in \cite{edmexp}
The experimental limit is
$|d_e| <  6.2 \times 10^{-27} e {\rm cm}$ at 95\% C.L..
The presence of a \cp-violating \zgg\ coupling can in principle 
be severely constrained from
this limit on the EDM of the electron \cite{sdr}. 
The effective interaction considered here can induce, at one loop, an EDM
for the electron. This calculation depends on a momentum cut-off.
It was shown in \cite{sdr} that with an assumed
cut-off of the order of 1 TeV, the experimental limit on the EDM
of the electron gives a limit
of about $10^{-3}$ on $\vert\lambda\vert$. The limit would be  
even more stringent, about $10^{-4}$, if the
cut-off is assumed to be of the order of a grand unification scale, viz., 
$10^{16}$
GeV. Thus, for values of $\lambda$ which are relevant for the experiments we
discuss in this paper, the induced electron dipole moment would be in conflict
with experiment.
However, the procedure for obtaining a
dipole moment from $\lambda$ at one loop in a non-renormalizable effective 
theory is not rigorous. Such a calculation does not take into account the 
dependence of $\lambda$ on $q^2$ values of $\gamma$ and $Z$.
Moreover, there is the possibility of cancellations
between the $\gamma ZZ$ and
$Z\gamma\gamma$ contributions to the electron edm, which is assumed to be
absent.  For these reasons, it would therefore be desirable to obtain a direct
experimental limit on $\lambda$ rather than an indirect one.

We should compare the
limits we discuss here with those that could be achieved in the process $e^+e^-
\rightarrow \gamma Z$ discussed in \cite{sdr}. The limits mentioned there are
an order of magnitude better for comparable linear collider parameters.
However, whereas we have taken into account top decay, 
\cite{sdr} did not take into account details of $Z$ decay, and it remains to be
seen how much the results in \cite{sdr} would be effected by $Z$ detection
efficiencies.

We have neglected $CP$ violation in the decay of the top quark. A complete
study should take this into account. 
It is quite conceivable that linear $e^+e^-$ colliders would achieve better
luminosities than anticipated here. In that case, there would be a
corresponding improvements in the limits we derive.

We thank the referee for pointing out an error in an equation in an earlier
version of the manuscript. One of us
(S.D.R.) thanks Debajyoti Choudhury and Rohini Godbole for helpful
correspondence.

\end{document}